# Spatial computing in structured spiking neural networks with a robotic embodiment

**Sergey A. Lobov[1,2,3]\*, Alexey N. Mikhaylov[1], Ekaterina S. Berdnikova[1], Valeri A. Makarov [1,4], Victor B. Kazantsev[1,2,3]**

[1]Lobachevsky State University of Nizhny Novgorod, Gagarin Ave. 23, 603950 Nizhny Novgorod, Russia

[2]Neuroscience and Cognitive Technology Laboratory, Center for Technologies in Robotics and Mechatronics Components, Innopolis University, Universitetskaya Str. 1, 420500 Innopolis, Russia

[3]Center For Neurotechnology and Machine Learning, Immanuel Kant Baltic Federal University, Kaliningrad, Russia

[4]Instituto de Matemática Interdisciplinar, F. de Ciencias Matemáticas, Universidad Complutense de Madrid, 28040 Madrid, Spain

**\* Correspondence:**
Corresponding Author
lobov@neuro.nnov.ru

## Abstract

One of the challenges of modern neuroscience is creating a "living computer" based on neural networks grown *in vitro*. Such an artificial device is supposed to perform neurocomputational tasks and interact with the environment when embodied in a robot. Recent studies have identified the most critical challenge, the search for a neural network architecture to implement associative learning. This work proposes a model of modular architecture with spiking neural networks connected by unidirectional couplings. We show that the model enables training a neuro-robot according to Pavlovian conditioning. The robot's performance in obstacle avoidance depends on the ratio of the weights in inter-network couplings. We show that besides STDP, critical factors for successful learning are synaptic and neuronal competitions. We use the recently discovered shortest path rule to implement the synaptic competition. This method is ready for experimental testing. Strong inhibitory couplings implement the neuronal competition in the subnetwork responsible for the unconditional response. Empirical testing of this approach requires a technique for growing neural networks with a given ratio of excitatory and inhibitory neurons not available yet. An alternative is building a hybrid system with *in vitro* neural networks coupled through hardware memristive connections.

## 1    Introduction

Since the end of the XX century, the concept of a "living computer" has been developing, including the cultivation of neuronal cultures on multi-electrode arrays (Pamies et al., 2014; Potter et al., 1997) and their implementation in neurorobots or neuroanimats . Neuroanimats are proposed to give neural networks the ability to learn in the context of interaction with the environment, as it happens with natural neural networks in the brain (Meyer and Wilson, 1991; Potter et al., 1997; Reger et al., 2000). However, despite certain advances in this area (Bakkum et al., 2008; Shahaf et al., 2008), associative learning in neural networks grown *in vitro* is still limited by the lack of a universal approach that works as well as learning algorithms for artificial neural networks (ANNs).

One possible reason for the failure of early attempts to achieve associative learning is the homogeneous structure of the network (Dauth et al., 2016; Pimashkin et al., 2013, 2016). Experimental approaches are now emerging for growing heterogeneous networks that include separate subnets connected by unidirectional links. The only thing missing is the algorithms for such a "living computer" based on neural networks grown *in vitro*. Earlier, we proposed an approach to explain the problems of learning in unstructured neural networks by the competition between different pathways conducting excitation to a neuron or set of neurons (Lobov et al., 2017a, 2017b). Recently, the possibility to structure the network geometry by directing axon growth was demonstrated experimentally (Forró et al., 2018; Gladkov et al., 2017; le Feber et al., 2015; Malishev et al., 2015; Pan et al., 2015), which opens a new venue to build network architectures *in vitro*.

Recently, in our model study, we proposed an associative learning approach based on "spatial computation" (Lobov et al., 2020b). The approach uses the "shortest path rule": On the network scale, STDP potentiates the shortest neural pathways and depresses alternative longer pathways. We use spiking neural networks (SNNs) and spike-timing dependent plasticity (STDP) because SNN, in particular based on the Izhikevich model, exhibit the entire neurocomputational spectrum of behavior (Izhikevich, 2003, 2004, 2007), and STDP is an experimentally proven form of Hebb's plasticity (Bi and Poo, 1998; Markram et al., 1997, 2011; Sjöström et al., 2001). So far, our approach has been illustrated at the scale of small neural circuits containing units of neurons. The question remained open: Is it possible to train a medium-scale network containing hundreds and thousands of neurons using the "spatial computation"? If we give an affirmative answer to this question, we thereby provide a possible algorithm for a "living computer".

Another important question that requires an answer according to the research results is the possibility of the subsequent implementation of heterogeneous networks *in vitro* with artificial elements based on the developed algorithms. The key elements of such networks are unidirectional connections between subnets with the property of synaptic plasticity. Memristors and memristive systems (Chua and Kang, 1976), which are implemented in the form of a simple CMOS-compatible nanostructure with a memory effect, are ideally suited for the role of such connection (Strukov et al., 2008). The first step in this direction has already been taken recently: commercial memristive devices with the effect of short-term plasticity are used to arrange communication between individual subnets *in vitro* and provide synchronous activity of target subnets under the control of the source subnet (Dias et al., 2021). In accordance with the general concept of memristive neurohybrid systems (Mikhaylov et al., 2020) and the first experimental results (Dias et al., 2021; Juzekaeva et al., 2019), it is the application of memristive devices and systems that will provide the necessary balance in terms of miniaturization, energy efficiency, and computational capabilities required for the hardware implementation of adaptive electronic interfaces between living neurons and their networks.

Earlier we formulated the basic principles of associative learning in SNN: (i) Hebbian learning (STDP); (ii) synaptic competition or competition of SNN inputs; (iii) neural competition or competition of SNN outputs (Lobov et al., 2020a, 2020b). The aim of the current work is to consistently implement these principles in a medium-scale SNN consisting of several subnets connected by unidirectional links.

## 2    Models and methods

To simulate the dynamics of a SNN, we adopt the approach described elsewhere (Lobov et al., 2017a). Briefly, the dynamics of a single neuron is given by (Izhikevich, 2003):

$$\frac{dv}{dt} = 0.04v^2 + 5v + 140 - u + I(t), \qquad (1)$$

$$\frac{du}{dt} = a(bv - u), \qquad (2)$$

where $v$ is the membrane potential, $u$ is the recovery variable, and $I(t)$ is the external driving current. If $v \geq 30$, then $v \leftarrow c$, $u \leftarrow u + d$, which corresponds to generation of a spike. We set $a = 0.02$, $b = 0.2$, $c = -65$, and $d = 8$. Then, the neuron is silent in the absence of the external drive and generates regular spikes under a constant stimulus, which is a typical behavior of cortical neurons (Izhikevich, 2003, 2004). The driving current is given by:

$$I(t) = \xi(t) + I_{syn}(t) + I_{stml}(t), \qquad (3)$$

where $\xi(t)$ is an uncorrelated zero-mean white Gaussian noise with variance $D$, $I_{syn}(t)$ is the synaptic current, and $I_{stml}(t)$ is the external stimulus. As a stimulus, we use a sequence of square electric pulses of the duration of 3 ms delivered at 10 Hz rate, with the amplitude sufficient to excite the neuron.

The synaptic current is the weighted sum of all synaptic inputs to the neuron:

$$I_{syn}(t) = \sum_j g_j w_j(t) y_j(t), \qquad (4)$$

where the sum is taken over all presynaptic neurons, $w_j$ is the strength of the synaptic coupling directed from neuron $j$, $g_j$ is the scaling factor, in this paper we set them equal to 20 or $-20$ (Lobov et al., 2017a) for excitatory and inhibitory neurons, respectively, and $y_j(t)$ describes the amount of neurotransmitters released by presynaptic neuron $j$.

To model the neurotransmitters, we use Tsodyks-Markram's model (Tsodyks et al., 1998) that accounts for short-term depression and facilitation. We use this model with the following parameters: the decay constant of postsynaptic currents $\tau_I = 10$ ms, the recovery time from synaptic depression $\tau_{rec} = 50$ ms, the time constant for facilitation $\tau_{facil} = 1$ s.

The dynamics of the synaptic weight $w_{ij}$ of coupling from an excitatory presynaptic neurons $j$ to a postsynaptic neuron $i$ is governed by the STDP with two local variables (Morrison et al., 2008; Song et al., 2000). Assuming that $\tau_{ij}$ is the time delay of spike transmission between neurons $j$ and $i$, a presynaptic spike fired at time $t_j$ and arriving to neuron $i$ at $t_j + \tau_{ij}$ induces a weight decrease proportional to the value of the postsynaptic trace $s_i$. Similarly, a postsynaptic spike at $t_i$ induces a weight potentiation proportional to the value of the presynaptic trace $s_j$. The weighting functions obey the multiplicative updating rule (Morrison et al., 2008; Song et al., 2000). Thus, the weight dynamics is given by:

$$\frac{ds_i}{dt} = -\frac{s_i}{\tau_S} + \sum_{t_i} \delta(t - t_i), \tag{5}$$

$$\frac{ds_j}{dt} = -\frac{s_j}{\tau_S} + \sum_{t_j} \delta(t - t_j - \tau_{ij}), \tag{6}$$

$$\frac{dw_{ij}}{dt} = \lambda \big[ (1 - w_{ij}) s_j \delta(t - t_i) - \alpha w_{ij} s_i \delta\big(t - t_j - \tau_{ij}\big) \big], \tag{7}$$

where $\tau_S = 10$ ms is the time constant of spiking traces, $\lambda = 0.001$ is the learning rate, and $\alpha = 5$ is the asymmetry parameter.

The modular SNN contained subnets, each of which included 500 Izhikevich neurons. By default, the ratio between excitatory and inhibitory neurons was 1:4. Connections between neurons were predominantly local. Subnets were connected by unidirectional connections.

We detected network bursts of spikes for each subnet as follows. In a time window of 50 ms, the total number of spikes generated by the subnet was counted. The value of 50 spikes was taken as the burst generation threshold; the time at the moment the threshold was exceeded was considered the burst start. When a burst was generated in the leading (presynaptic) subnet in an interval of 100 ms, the presence of the beginning of generation of a synchronous burst in the slave (postsynaptic) subnet was checked. The network burst transmission percentage was calculated based on the ratio of the number of bursts in the postsynaptic subnet, synchronous with the presynaptic, to the total number of bursts in the presynaptic subnet.

We implemented the SNN model as custom software NeuroNet developed in QT C++ environment. The app supports SNNs with up to $10^4$ neurons. On an Intel® CoreTM i3 processor, the simulation can be performed in real time for a SNN with tens of neurons.

## 3    Results

### 3.1    Self-reinforcing effect of connections in neural circuits

In model studies, the efficiency of connections between neurons is usually understood as their weight, $w$, which determines the synaptic current arising in postsynaptic neuron when a spike is generated by presynaptic one (4). However, it is impossible to measure $w$ under experimental conditions; therefore, the efficiency is estimated indirectly by the amplitude of the postsynaptic potential or by the number of spikes "transmitted" from one neuron to another. Likewise, it is possible to determine the effectiveness of connections between subnets in experiments *in vitro* (Pan et al., 2014, 2015; Pigareva et al., 2021). Let us consider in the model both cases - with separate neurons (Fig. 1A) and subnets (Fig. 1B). For the efficiency of connections, $P$, here we will take the percentage of activity (individual spikes or their bursts) that arose in the second, postsynaptic neuron (subnet) a short time after registration of activity in the presynaptic neuron (subnet). The dependence of the efficiency of connections $P$ on $w$ has a pronounced sigmoid character both in the case of individual neurons (Fig. 1C) and in the case of subnets (Fig. 1D represents the dependence of $P$ on $W$ - the average weight of inter-subnet connections). Note, that a nonlinear dependence of the slave subnet activity on the master one is also observed in experiments (Pan et al., 2014, 2015).

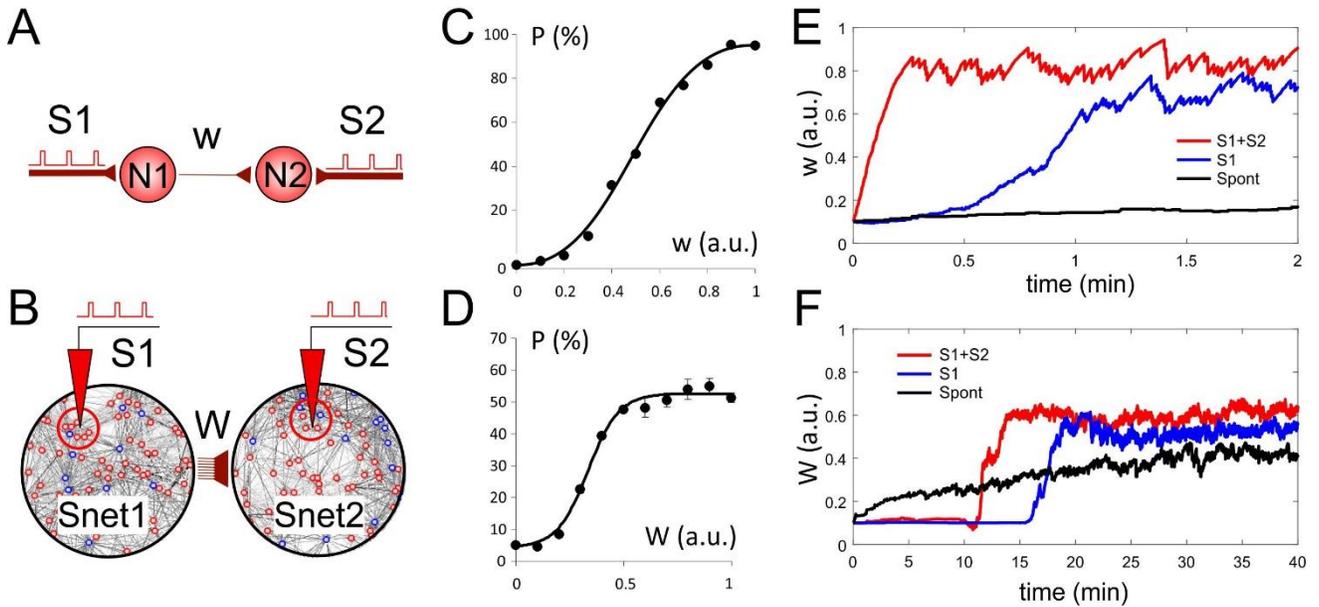

**Figure 1**: Self-reinforcing effect of connections in network structures: A, B) Scheme of paired stimulation of neurons (subnets) connected by unidirectional connections; C) Percentage of synchronous activity of connected neurons depending on the weight of the connection; D) Percentage of synchronous activity of connected subnets depending on the average weight of connections between subnets (the number of connections $N = 10$); E, F) Dynamics of changes in weights under conditions of spontaneous activity (Spont), stimulation of a presynaptic neuron or subnet (S1), paired stimulation (S1 + S2).

The presence of STDP in unidirectional connections can lead to a self-reinforcing effect of – i.e., to potentiation in the case of "effective" transfer of spikes through a connection. This effect is observed especially quickly when the STDP protocol of paired stimulation is applied, in which the presynaptic neuron is first stimulated, and after a short period of time (10 ms) - the postsynaptic neuron (Fig. 1E, S1 + S2) However, potentiation also occurs in the presence of stimulation of only the presynaptic neuron (fig. 1E, S1). And even with the only spontaneous activity, we can observe slow potentiation (Fig. 1E, Spont). This is because the presynaptic neuron sometimes excites the postsynaptic one, and according to the STDP rule, this leads to the potentiation. In a circuit with neural subnets, these patterns also appear. However, they develop much more slowly and with much smaller differences between evoked and spontaneous activity (Fig. 1F).

STDP-based self-reinforcing effect may underlie the formation of neural structures with cyclic activity and, possibly, central rhythm generators, CPG. Consider a system consisting of four subnets closed by unidirectional connections (Fig. 2 A). With low weights of interconnection and/or an insufficient number of them, the activity of subnets is practically uncorrelated (Fig. 2 B). Due to the rather high probability of temporal overlap of packs in the absence of connections, our method shows the average percentage of packets passing $P = 27\%$, which can be considered as the basic level. In the presence of a sufficient number of connections between subnets, the self-reinforcement effect leads to their potentiation and the emergence of circulating activity (Fig. 2 C). In this case, neural activity is transmitted from one subnet to another, and complete cycles can be repeated. The value of $P$ can reach 80% or more. Note, that the possibility of this effect occurring is determined by the number of connections $N_W$ between subnets (Figure 2 D). E.g., in the model with $N_W < 4$, no circulating activity was observed. In the presence of bidirectional connections, self-reinforcement of pathways is also observed - after a certain time of STDP rearrangements, there is an increase in connections that provide the circulation of activity in the direction either clockwise or counterclockwise.

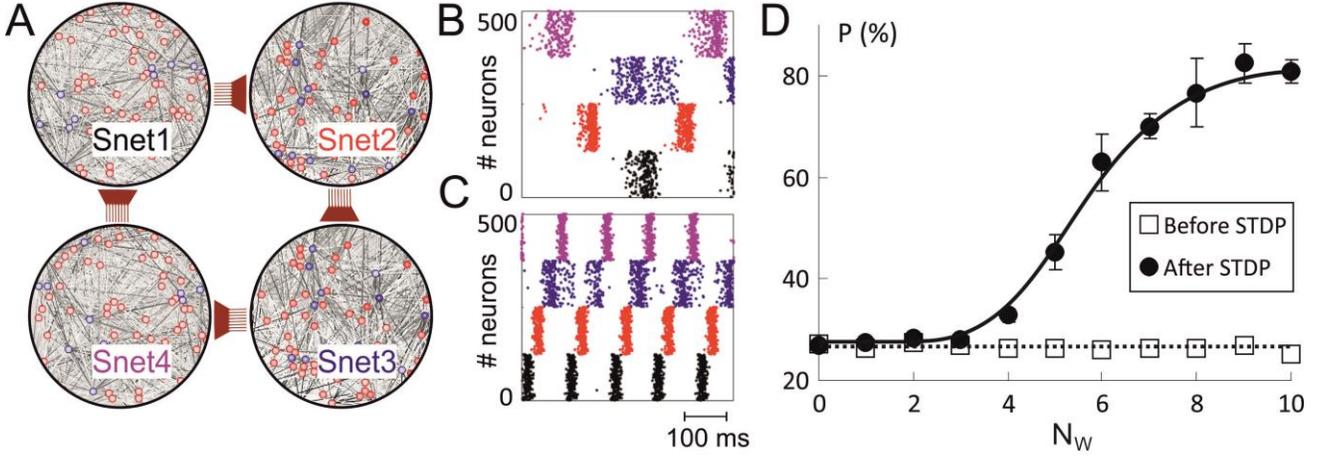

**Figure 2**: Cyclic structures and circulating activity. A) Network architecture, including subnets connected by unidirectional links; B, C) An example of raster diagrams before and after learning, demonstrating the occurrence of neural activity circulating through the network. D) Dependence of the efficiency of passing busts from one subnet to another on the number of connections between subnets before and after learning.

### 3.2 The shortest pathway rule

One can try to implement the simplest associative links on a two-block architecture (Fig. 1). In this case, the activity of the postsynaptic neuron (subnet) can be considered as the output response of the system. Accordingly, in terms of the Pavlovian conditioning, the stimulation of the presynaptic neuron (subnet) will be unconditional signal, and the stimulation of the first neuron will be conditional. However, here we are faced with a number of problems: spontaneous (without stimuli) strengthening of connections, strengthening of connections when only a conditioned stimulus is given (no association), and finally, the absence of a mechanism for removing an association when it is irrelevant (for example, when the conditional stimulus is not supported by an unconditional one). Thus, when trying to implement associative learning, the main problem is not the strengthening of the connections that carry out the association, but with the weakening of the connections that are not involved in the association of stimuli.

We previously described an effect called the shortest pathway rule and proposed a simple neural circuit with associative learning based on it (Lobov et al., 2020b). Consider this phenomenon on neural networks connected by unidirectional connections (Fig. 3). In the presence of two alternative ways of conducting excitation (W1 and W2, Fig. 3A), the shortest path (W1, Fig. 3B) is potentiated. In this case, synapses involved in the transmission of excitation along a longer pathway are depressed (W2, Fig. 3B). Further we will use both properties of the rule: potentiation and depression. Since the quality of training of the studied network architectures is ultimately determined by both properties, we introduce the coefficient of the quality of learning:

$$Q = \frac{2W_{pot}}{W_{pot}+W_{dep}} - 1, \qquad (8)$$

where $W_{pot}$ is the average value of connections between subnets that should be potentiated in the learning process, $W_{dep}$ is the average value of connections that should be depressed. In cases where learning is poor, $Q$ has a value close to zero, with "wrong" learning the value of $Q$ is negative. We will conventionally assume that neural architectures with Q > 0.5 are properly trained.

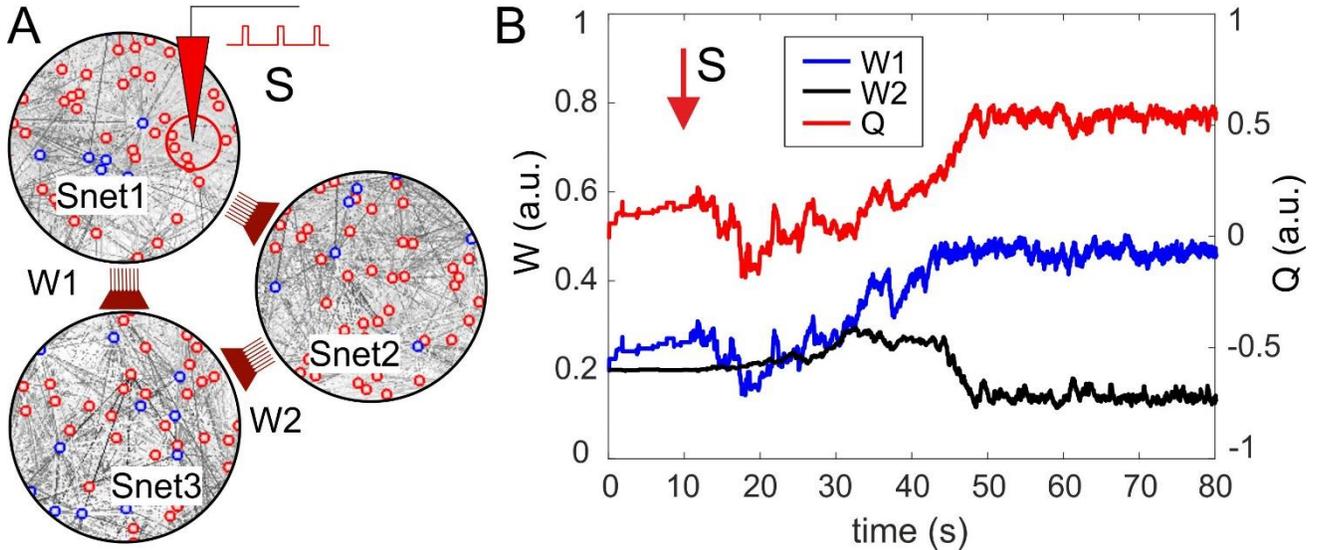

**Figure 3**: The shortest pathway rule for subnets structure. A) The network architecture. B) Dynamics of weights and the coefficient of the quality of learning

### 3.3   Synaptic competition in SNN

The shortest pathway rule can be used to implement synaptic competition, which solves problems with uncontrolled growth of the efficiency of connections and allows us to remove associations if they are irrelevant. Consider network architectures with two input elements and one output element (Fig. 4). In terms of Pavlovian conditioning, this architecture allows us to model a situation with two possible conditional stimuli CS and one unconditional stimulus US. During the learning process, the output element (N3, Fig. 4A; Snet3, Fig. 4D) will change the weights of the incoming connections depending on the correlation of its activity with the activities of the input elements. In the implemented protocol (Fig. 4B, Fig. 4E), "incorrect" pre-training occurs first when combining US with CS1. Then the main stage of training is carried out, in which US is combined with CS2. Such a protocol allows us to analyze not only the ability of the system to form associative links, but also the ability to retrain when changing external conditions (in our example, replacing CS1 with CS2).

Bidirectional connections between input elements ($w_c$, Fig. 4A; $W_C$ in Fig. 4D) are a key element of synaptic competition. They close the alternative long path when one of the input elements is jointly activated. As a result, along with the strengthening of the currently relevant associative connection, a weakening of the irrelevant association occurs and, accordingly, the learning coefficient $Q$ increases (Fig. 4B, 4E). The results of simulations with different weights of the competition connection show a certain range for which optimal learning is observed (Fig.4C, 4F, pink area). Combining two input subnets (Snet1 and Snet2) into one in our simulations also provided good learning with $Q = 0.63 \pm 0.11$ ($n = 6$). Comparing the learning results obtained in a chain of individual neurons and a network with subnets, it can be noted that in the second case, learning is slower (Fig. 4B vs. 4E) and with lower quality (Fig. 4C vs. 4F). This is due to the more complex neural dynamics of subnets and the presence of stimulus-induced synaptic rearrangements not only between subnets, but also within subnets.

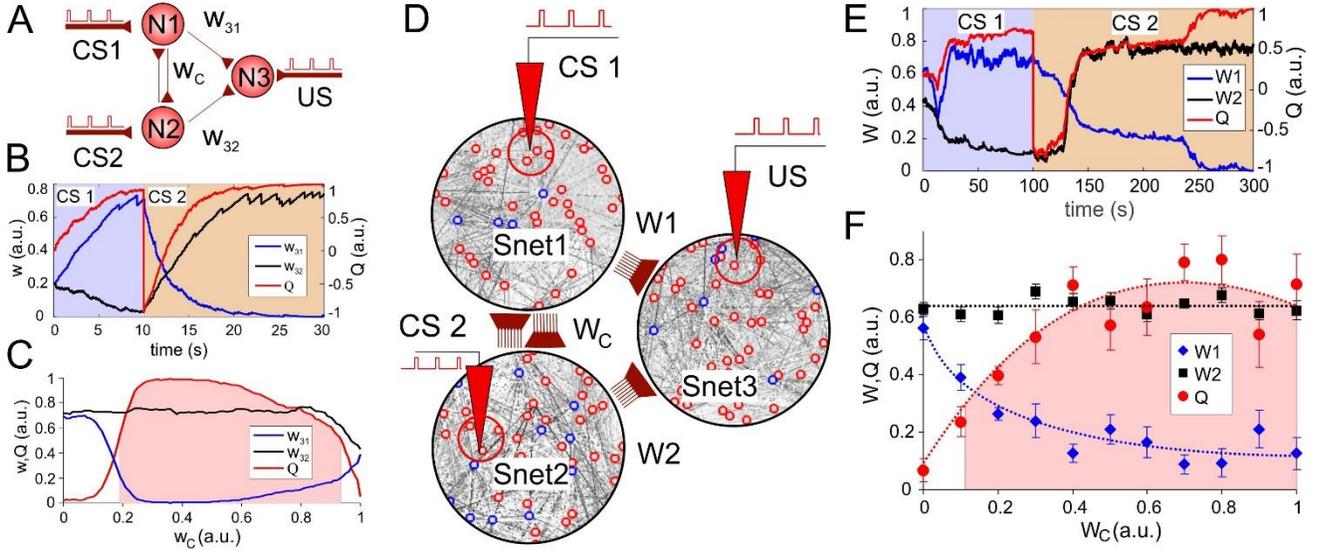

**Figure 4**: Associative learning and synaptic competition.

### 3.4 Neural competition of SNN outputs

A neural network can have not only several possible input (conditional) stimuli, but also several output (unconditional) ones. Therefore, for effective learning, it is necessary to implement not only competition between input neural elements (synaptic competition) but also competition between output elements – let us call it neuronal. Consider a circuit with one input and two output (Fig. 5A, C). In terms of Pavlovian conditioning, this architecture allows one to model a situation with one possible conditioned stimulus CS and two unconditioned stimuli CS. Here, as before (Lobov et al., 2020b, 2020a), to implement neural competition, we will use lateral inhibition, which suppresses the activity of neighboring neuronal elements upon strong activation ("win") of one of the elements. As a result of this process, only the winning element undergoes learning in the form of association. Since in this case we do not use the mechanism for removing irrelevant associations, the training protocol does not contain a phase of "wrong" pre-training. In the case of subnets, we combined two output networks into one, bearing in mind the impossibility of forming exclusively inhibitory connections between subnets in experimental conditions.

Let us investigate how the parameters of the inhibitory elements determine the quality of learning. As such parameters, we chose the weight of inhibitory connections ($w_I$ in the case of circuits from individual neurons and $W_I$ in the case of a system from subnets) and the decay time of inhibitory postsynaptic current $\tau_I$ in the Tsodyks-Markram's model (Tsodyks et al., 1998). The simulation results (Fig. 5B, D) show that in both systems under study, learning fails for $w_I$, $W_I < 0.1$ and $\tau_I < 40$ ms. Thus, lateral inhibition is a necessary element of learning. As with the case with two inputs and one output, the learning quality is better with individual neuron circuits (Fig. 5B, $Q_{max} = 0.8$) than with the system of subnets (Fig. 5B, $Q_{max} = 0.7$). This is due to the presence of incomplete suppression of competing neurons in a large network even at extremely high values of the parameters of inhibitory connections.

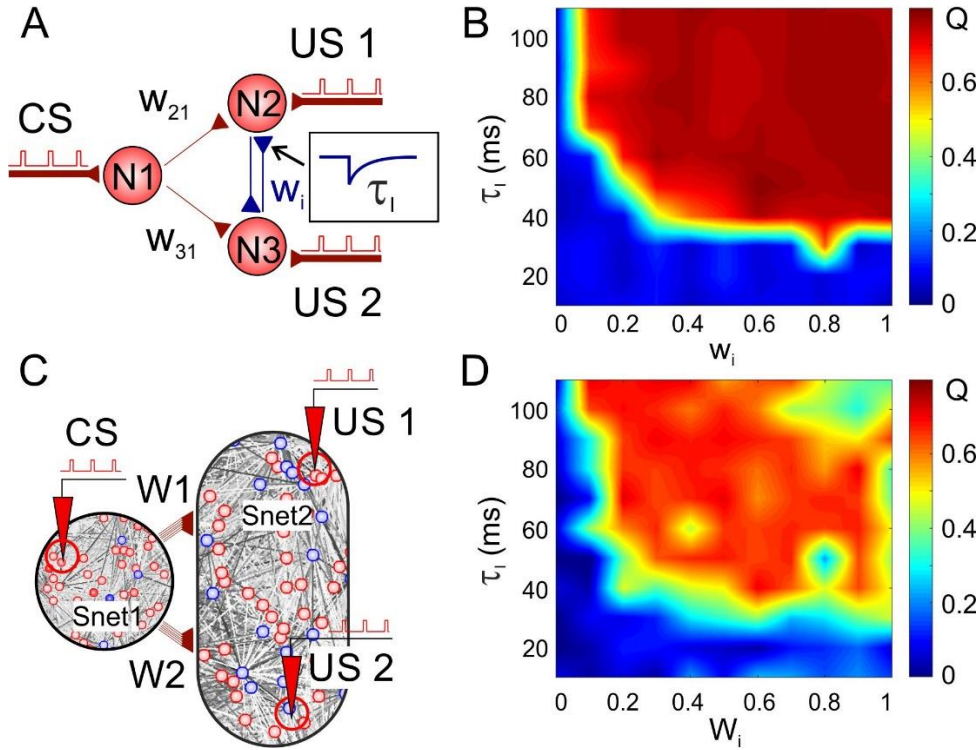

**Figure 5**: Associative learning and neural competition.

### 3.5 Associative learning with robotic embodiment

Earlier, we proposed a scheme for mapping activity of neural circuit to "behavior" of mobile robot training in the context of obstacle avoidance problem (Lobov et al., 2020b). In particular, a LEGO robot was used, with two touch sensors and two sonars (Fig. 6A, B). The activity of the motor neuron-pacemaker (N7 Fig. 6A, C) was translated into the command of rotation of the right and left motors, which led to the movement of the robot forward. The signal from touch sensors served as an unconditioned stimulus - upon stimulation of the corresponding neurons (N3, N4, Fig. 6C), excitation was transmitted to motoneurons that brake the wheels. As a result of such an unconditional reaction, the robot could a priori avoid obstacles upon contact with them. The signal from the sonars served as a conditional stimulus - when one of the CS neurons (N1 or N2, Fig. 6C) was activated, simultaneously with the activation of one of the US neurons (N3 or N4, Fig. 6C), the corresponding connections were strengthened, and the robot learned to go around obstacles in advance without contacting them. Training the robot could mimic Pavlovian conditioning (when stimuli were given regularly from the left and right sides, Fig. 6B), or operant learning (when the robot moved in a free mode, receiving stimulation from objects that it encountered on its way).

We have implemented a similar robotic embodiment in the case of a subnetted neural system (Fig. 6D). In this case, we used two subnets - for stimulation with conditional stimuli (Snet 1, Fig. 6D) and unconditional stimuli (Snet 2, Fig. 6D). At the same time, the internal connections in the Snet 1 subnet provided synaptic competition for interconnections $W_P$ and $W_D$ projected onto subnet Snet 2. In turn, the inhibitory neurons of the Snet 2 provided neural competition - when one of the zones of the Snet 2 was excited, the other zone was inhibited. Due to the duration of the training process and the poorer quality, in the case of a neural system consisting of subnets, training can be carried out only in the mode of simulating a Pavlovian conditioning (the dynamics of learning is shown in Fig. 6E). Testing robots driven by SNNs with different learning quality values shows relationship between $Q$ and the robot's "behavior" (Fig. 6F). Note, that the dependence of the number of collisions on $Q$ fits well with an exponential function, both for a neural system with subnets (Fig. 6F, Subnets) and individual neurons (Fig. 6F, Neurons). The higher the quality of learning, the fewer collisions are registered during the test time (10 min), which really allows us to call $Q$ the quality of learning. Since the dynamics of a complex neural system negatively affects the robot's behavior, the number of collisions in the case of using a simple neural circuit is lower over the entire range of $Q$ values.

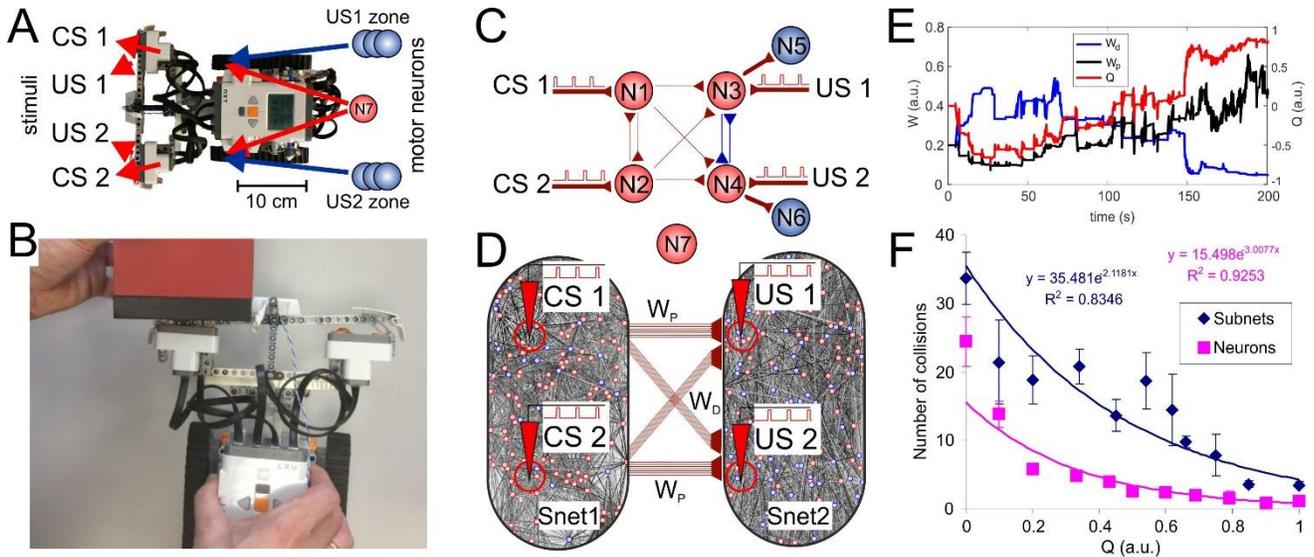

**Figure 6**: Neurobot and the relationship of the network and behavioral level. A) Mapping of the sensory and motor neurons in the mobile LEGO robot. B) Training the robot, in which the touch sensor and the ultrasonic sensor are simultaneously triggered. C) Neural circuit with associative learning. D) Modular network, consisting of subnets connected by unidirectional connections ($W_P$ and $W_D$). F) The dynamics of weights of unidirectional connections ($W_P$ and $W_D$) and the coefficient of quality of learning (Q) in the learning process. D) Dependence of the number of collisions of the robot with obstacles on the coefficient of the quality of learning in case of the neural circuit ("Neurons") and the modular network ("Subnets").

## Conclusions

In this paper, we investigated the architectures of modular SNNs with subnets connected by unidirectional connections. First, we studied a two-block architecture, consisting of two identical subnets. Simulating has shown that the presence of STDP in unidirectional connections can lead to the effect of self-reinforcement of synapsis with efficient spike transfer. This effect is observed with the use of stimulation of the presynaptic network, with paired stimulation according to the STDP protocol, as well as in the presence of spontaneous activity caused by neural noise. It is shown that a consequence of the strengthening of connections between subnets is an increase in the number of spike busts passing from one subnet to another, while the dependence is sigmoidal. An attempt to use the two-block architecture in associative learning has identified a number of factors that have a negative impact on learning. In particular, the main problem was not the strengthening of the connections that carry out the association, but the insufficient weakening of the connections that are not involved in the association of stimuli.

Next, we studied the possibility of an architecture with two input (presynaptic) subnets and one postsynaptic subnetwork. Such SNN can implement associative learning with two conditional signals and unconditional one. It is important to ensure competition between converging interconnections here. Since in the case of a simple neural chain, converging connections are synapses of one neuron, we called this phenomenon synaptic competition. In the current work, as before for neural circuits (Lobov et al., 2020b), to implement synaptic competition, we used the "shortest path rule": On the network scale, STDP potentiates the shortest neural pathways and depresses alternative longer pathways. It is important to note that this rule (Fig. 3), as well as the architecture with converging connections (Fig. 4), can already be tested experimentally *in vitro*.

The next step involved study the architecture with one input and two outputs. Such SNN is able to implement associative learning with one conditional signal and two unconditional ones. It is important here to ensure competition between diverging connections. Earlier, we called this phenomenon neural competition, meaning that postsynaptic neurons compete with each other, suppressing their neighbors upon excitation. In the current work, as before (Lobov et al., 2020b, 2020a), we implemented neuronal competition using lateral inhibition. This approach is still impossible to test experimentally, since a technique for growing neural networks with specified parameters for excitatory and inhibitory elements has not yet been proposed.

Finally, we have proposed the network architecture capable of handling two conditional and two unconditional signals and providing two associative links. Associative learning is demonstrated using a neurorobot. In the interface scheme, the signal from the robot's sensors is fed to certain sections of the network, providing the possibility of synchronous activity in a pair of sections. Before training, the robot can only avoid obstacles when it collides with them. When learning, which consists in presenting the robot with an obstacle in one of the sides, the conditioned stimulus (CS), mediated by the ultrasonic sensor, is associated with the unconditioned stimulus (US), mediated by the touch sensor. As a result of training, the robot can go around obstacles without contacting them. To characterize the quality of training, we proposed to use the coefficient of the learning quality Q, based on the ratio of the weights of connections between subnets. During training, Q can increase up to its maximum value. Experiments with a robot controlled by SNNs with different learning quality revealed an exponential relationship between the Q and the number of errors, i.e. collision with obstacles.

The results of this study open up a long-term perspective related to the development of therapeutic strategies and neurotechnologies based on direct modulation of neural electrical activity in the treatment of neurological disorders (monitor-compute-actuate paradigm). In this case, the developed algorithms for learning / stimulating brain subnets and electrically controlled connections between such subnets will be useful. Adaptive memristive connections created artificially will provide the necessary degree of freedom in setting the parameters of connections (their weights and their number) for the occurrence of correlated activity of living subnets. Further development of this technology may be associated with the integration of control circuits, memristive and microelectrode arrays on a single CMOS chip with subnets spatially ordered by microfluidics (Mikhaylov et al., 2020).

**Conflict of Interest**

The authors declare no conflict of interest.

**Acknowledgments**

This work was supported by the Russian Science Foundation (grant No. 21-12-00246, SNN simulation; grant No. 19-12-00394, experiments with the neurorobot) by the Russian Foundation for Basic Research (grant No. 20-01-00368, spatial neurocomputation concept; grant No. 18-29-10068, network synchronization; grant No. 18-29-23001, emulation of synaptic plasticity by using memristive devices and systems).